\documentclass[reprint,
 amsmath,amssymb,
 aps,
]{revtex4-2}

\usepackage{graphicx}
\usepackage{dcolumn}
\usepackage{bm}

\usepackage{ulem}
\usepackage{xcolor}

\begin{document}

\preprint{APS/123-QED}

\title{Machine Learning Phase Field Reconstruction in a Bose-Einstein Condensate}

\author{Jackson Lee}
\author{Andrew J Millis}%
 \altaffiliation[Also at ]{Center for Computational Quantum Physics, Flatiron Institute }
\affiliation{%
 Department of Physics, Columbia University
}%

\date{\today}

\begin{abstract}
A basic challenge in experimental physics is the extraction of information related to variables that are not directly measured. The challenge is particularly severe in quantum systems where one may be interested in correlations of operators that are not diagonal in the measurement basis.  In this paper we take a step towards addressing this issue in the context of Boson superfluids, where standard in-situ imaging yields only the spatially resolved density, leaving the phase field—crucial for identifying topological defects such as vortices and confirming superfluidity—indirectly encoded. Previous work has shown that the location of vortices in the phase field may be detected, but has not solved the problems of fully reconstructing the phase or identifying the charge (vortex vs. antivortex). This paper shows that a combination of a deep machine learning (ML) model and classical computer vision post-processing steps can address this gap. We use realistic snapshots of the thermal state of a two-dimensional BEC in a harmonic trap using synthetic data obtained from projected Gross-Pitaevskii equation simulations to train a U-Net-based architecture to infer the absolute values of the phase field gradients from an observed density field, and then employ a separate ML model to locate the positions of the vortex cores  and a post-processing graphical analysis to determine with high accuracy the phase field, including the quantized charge of each vortex.

\end{abstract}

\maketitle


\section{\label{sec:level1}Introduction}

Over the past two decades, advances in cooling and trapping techniques have enabled unprecedented experimental control over quantum systems, enabling the creation of remarkable quantum states of matter \cite{gorlitz_realization_2001, abo-shaeer_observation_2001, klaus_observation_2022, bottcher_transient_2019, bigagli_observation_2024, yuan_extreme_2025, dalibard2024magnetic, chomaz_dipolar_2022, schindewolf_few-_2025, basov_towards_2017, bloch_strongly_2022, khajetoorians_creating_2019, paschen_quantum_2020, defenu_out--equilibrium_2024}.  However, inferring quantum correlations remains an experimental challenge. Available imaging techniques mainly work by projection of the quantum state onto a particular (typically density) basis \cite{adamson_improving_2010, ketterle_making_1999, kosaka_spin_2009, parsons_site-resolved_2016}. Impressive advances provide very fine-grained  information on atom positions, but obtaining information about other correlations, including phase and spin variables, that are conjugate to (non-commuting with) the more straightforwardly measured variables, remains challenging.

One possible route is via 'Hamiltonian learning' \cite{Tucker24,Liu25,Guo25}, in which a many-body Hamiltonian is inferred from short time dynamics measurements and the inferred Hamiltonian is then used to compute the desired correlations. However, it is desirable to have a method that learns correlations directly from "snapshot" data obtained by projection onto one basis at a given time, and does not require measurements of dynamics. An argument that such a reconstruction may be possible in principle follows from the logic underpinning the Hohenberg-Kohn theorem \cite{Hohenberg64} of electronic structure. This theorem says that the ground state wave function of a quantum system is fully determined by the particle density (space dependent one particle density matrix), if the interparticle interaction is specified. If such a relationship exists, then it may be possible to use AI methods to learn the functional relationship and predict other correlations from data. While many papers have attempted to learn the Hohenberg-Kohn density functional of electronic structure theory \cite{Burke12,Nelson18,Moreno20,Bai22,Kumar23,Cronin25}, the problem of reconstructing quantum correlations directly from measured densities has received less attention. 

In this paper we apply this idea in the context of Bose–Einstein condensates (BECs). Formed by cooling a dilute gas of bosons to ultralow temperatures, BECs exhibit striking collective phenomena such as long-range coherence, superfluidity, and quantized circulation \cite{dalfovo_theory_1999}, all of which relate to the correlations of the quantum mechanical wave function phase, which is conjugate to the particle density. Typical  time-of-flight (TOF) expansion or in-situ imaging experiments extract the density profile with high accuracy, but in these experiments the phase of the wave function is not directly observed \cite{ketterle_making_1999, bloch_many-body_2008, dalfovo_theory_1999}. Experimental efforts to extract phase information rely on interference between different parts of a condensate \cite{hagley_measurement_1999}, or between different condensates \cite{zawadzki_spatial_2010, stock_observation_nodate}. Theoretical approaches, typically relying on TOF information, have also been developed. These techniques rely on using many repetitions of an experimental setup \cite{sunami_detecting_2025, meiser_reconstruction_2005}, several measurements in-situ and in the course of a TOF expansion \cite{tan_wave-function_2003, ziv_phase_2024}, or TOF measurements with knowledge of in-situ support constraints \cite{kosior_condensate_2014}. These approaches typically require expensive iterative schemes, have limited spatial resolution, and may require several snapshots in a single run, which is difficult when the measurements are destructive.

A particularly important feature of the phase field of a BEC is its topological defects— vortices and antivortices—which appear as density depletions around which the phase winds by an integer multiple of  $2\pi$ \cite{dalfovo_theory_1999}. The configuration, dynamics, and interactions of these defects play a central role in two-dimensional superfluids, particularly in the context of vortex–antivortex unbinding across the Kosterlitz–Thouless transition \cite{berezinskii1972destruction, kosterlitz_ordering_1973}. As a result, experimental access to the condensate phase is essential for characterizing superfluid order and identifying distinct phases of matter. Recent work has applied machine learning and computer vision techniques to identify vortex cores within noisy density images with high accuracy \cite{ye_u-net_2023, metz_deep-learning-based_2021}. However, these methods generally stop at locating vortex positions: they do not reconstruct the full phase field, nor distinguish vortices from antivortices (unless given the corresponding phase profile, as in \cite{metz_deep-learning-based_2021}).

In this paper, we present a machine-learning-based approach that successfully reconstructs the phase field of a superfluid from measurements of the density. Attempts to use a neural network to learn the full (signed) phase field were not successful. The method proposed here is a two-stage approach in which a U-Net convolutional neural network architecture, known for its power in image-to-image translation, is used to determine the magnitude of the gradients of the phase field. The signed phase gradients are then determined using a classical post-processing method. The U-Net was trained on synthetic data obtained from numerical solutions of the Projected Gross Pitaevskii Equation (PGPE) (representing the  long-wavelength modes of the system) combined with a uniformly distributed random density (modelling the incoherent, short-wavelength modes). We note that in the absence of the  incoherent background the vortex cores are points of vanishing density, so vortices may easily be identified; the presence of the incoherent background means the vortex cores are simply regions of lower than average density,  making the vortex identification problem  significantly more challenging. 

The U-Net architecture identifies lines where $\partial_{x}\phi=0$ and another set of lines where $\partial_y\phi=0$, where $\phi$ is the phase; these lines bound domains where a component of $\nabla\phi$ has a definite sign. The remaining problem is to identify the signs by "coloring" the  domains consistently; this problem is complicated by the fact that  the ML-predicted borders are returned as probabilities, and are not guaranteed to be always correct. We introduce a classical post-processing that finds  a configuration of the sign of $\partial_x\phi$ that maximally respects the ML-predicted borders.

The rest of this paper is organized as follows. Section ~\ref{sec:level1} presents the needed background, including the theoretical definition of the Bose condensate problem and the PGPE approach to generating synthetic data, as well as general features of the machine learning methodology. Section ~\ref{sec:phasereconstruction} shows representative results. Section ~\ref{sec:algorithm} presents the essential features of the algorithm.  Section ~\ref{sec:conclusion} presents a summary, conclusions and outline of future directions. Technical aspects of the algorithm and calculations are presented in the Appendices.

\section{\label{sec:level1}Background}

\subsection{Non-zero Temperature Bose Gases}
We consider a two-dimensional (\textbf{2D}) Bose fluid in thermal equilibrium at a nonzero temperature but at zero magnetic field. The Bose fluid is characterized by a complex  field $\Psi$ representing the long-wavelength fluctuations and an additional field  representing the short-wavelength fluctuations  (loosely, uncondensed bosons). The long-wavelength fluctuations may be written as an amplitude and phase $\Psi(r,t)=\left|\Psi(r,t)\right|e^{i\phi(r,t)}$; typical configurations involve nontrivial phase configurations including  both vortices and antivortices. The short-wavelength fluctuations are represented by a density field $\rho_{inc}$. Inclusion of the incoherent background complicates the task of learning the phase: in the absence of the background, vortices and antivortices are characterized by a vanishing of particle density at the vortex core, allowing for easy identification.

Our goal is to determine the phase field from ``snapshot'' measurements of the density $\left|\Psi(r,t)\right|^2+\rho_{inc}$, by training a neural network on synthetic data.  The density to phase reconstruction problem is fundamentally ambiguous due to the combination of time reversal ($\Psi \to \Psi^*$) and global $U(1)$ ($\Psi \to \Psi e^{i\alpha}$) symmetries  of the Hamiltonian, which preserve the density while changing the phase.  In this paper, we avoid these issues by  predicting the \emph{gradient} of the phase, $\nabla\phi$,  up to a global sign. $\nabla\phi$ has the physical interpretation of the velocity field \cite{dalfovo_theory_1999}.

To generate synthetic data representing ``snapshots'' of the long-wavelength fluctuations we employ the Projected Gross-Pitaevskii Equation (PGPE), which has been used to study Bose fields and observe vortex unbinding at finite temperatures \cite{davis_simulations_2001, blakie_projected_2005, simula_thermal_2006, foster_vortex_2010}. In the PGPE, the \textit{classical field}, which consists of the long-wavelength, highly occupied modes, is evolved separately from the the \textit{incoherent background}, which consists of the short-wavelength, sparsely occupied modes. The classical field $\Psi(\mathbf{r}, t)$ is time-evolved via the PGPE:

\begin{equation}
i \hbar \frac{\partial \Psi}{\partial t} = \left( -\frac{\hbar^2 \nabla^2}{2m} + V_{\text{trap}}(\mathbf{r})\right) \Psi + \mathcal{P}\{g|\Psi|^2\Psi\}
\end{equation}
where $V_{\text{trap}}(\mathbf{r})$ is the trapping potential, $g$ is the interparticle interaction constant, $m$ is the atomic mass, and the projector $\mathcal{P}$ projects onto the long-wavelength modes, which are taken to be eigenstates of the trapping potential. Here, we utilize a radially symmetric trapping potential $V(x, y) = \frac{1}{2}m\omega^2(x^2 + y^2)$.  Time evolution of the PGPE corresponds to taking samples from a microcanonical ensemble. Here, we generate samples by allowing a random initial state to evolve to equilibrium, before taking samples at fixed time intervals (see appendix \ref{app:PGPE Simulations} for more details). 

The fact that we can use a classical field to represent the long-wavelength modes can be seen by the fact that those modes are highly occupied, with occupations $n_k\gg 1$.  Equivalently, the energy cutoff enforced by the projection introduces a coarse-graining, such that each coarse-grained region contains many particles and exhibits small relative phase fluctuations. Thus, in this regime, the field operator can be accurately approximated by a classical complex field with a well-defined amplitude and phase. A vortex is defined via a $2\pi$ winding of the phase of $\Psi$, and corresponds to regions where the c-field density $|\Psi|^2$ drops to 0 \cite{dalfovo_theory_1999}. For any closed contour on the 2D surface, we have:
\begin{equation}
\Gamma = \oint_C \nabla \phi \cdot d\mathbf{l} = 2\pi k
\label{quantization_eq}
\end{equation}

Here, $k$ must be an integer due to the single-valued nature of the wave function. A contour enclosing a vortex has $k = 1$, while a contour enclosing an antivortex has $k = -1$. More generally, a contour enclosing $n$ vortices and $m$ antivortices has $\Gamma = 2\pi(n - m)$. In this work, we detect ground truth vortices in the phase fields by computing the circulation around groups of 4 pixels.  

To model the incoherent part, the full PGPE formalism requires an estimate of the temperature and chemical potential from the properties of the equilibrium classical field; here, noting that the incoherent density is often constant in the spatial region of the classical field \cite{blakie_dynamics_2008}, we simply model the incoherent density as a background density. Thus, for sample $i$, we have a density field $|\Psi_i(\mathbf{r})|^2 + \rho_{inc}^{(i)}$, where the phase field is defined by $\phi = arg(\Psi_i(\mathbf{r}))$, and where $\rho_{inc}^{(i)}$ is the background density, chosen to be of the order of the average central density of the classical field, with the average taken across samples of the same energy. During training, however, the training examples are given a random background density, in order to help the model generalize to energies outside the training set.

\subsection{Network training}
We train our algorithm on snapshots generated at dimensionless energy (see appendices) $E = 23$, using a training set of 1125 density-phase pairs, and a validation set of 375 pairs used to tune hyperparameters for both the deep learning and classical portions of the algorithm. We then applied the algorithm to a test set of 2000 densities, consisting of snapshots with energies $E \in[13, 32]$. For each example in the test set, the model predicts both the value of $\nabla \phi$ up to a global sign, as well as the positions and charges of vortices up to a global sign, given only a density snapshot and the trapping potential (which is the same for every example) as inputs.

\section{Phase and Vortex Reconstruction from Density \label{sec:phasereconstruction}}

\begin{figure*}[t]
  \centering
  \includegraphics[width=\textwidth]{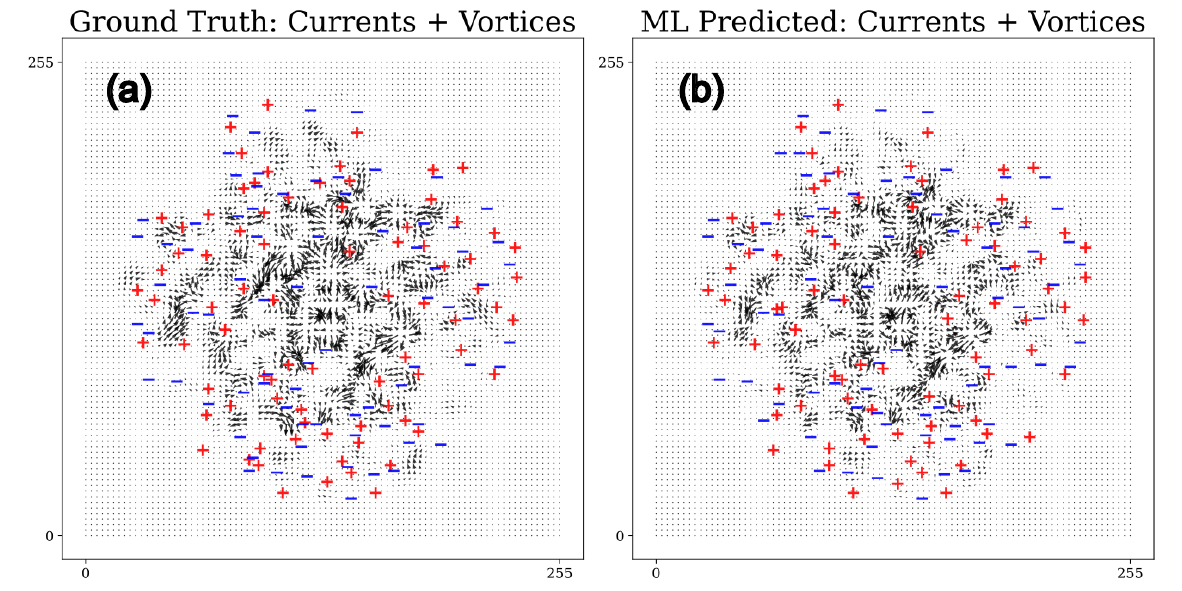}
  \caption{\label{fig:current_predictions_with_background} A representative example of an ML prediction on a test set example at the same energy as the training energy. Note that while our algorithm predicts the gradient of the phase field, $\nabla \phi$, here we visualize the current, $\rho\nabla\phi$, in order to create visually informative arrow plots (plotting $\nabla \phi$ causes exploding gradients near vortices). (a) A ground truth example of the current field drawn from the test set, at the same energy that the ML model was trained on. Ground truth vortices (antivortices) are denoted by +(-) markers. (b) ML prediction of the phase field, $\nabla\tilde{\phi}$, shown here as the current field $\rho\nabla\tilde{\phi}$, given only the ground truth density $\rho$ with a uniform thermal background. The ML predicted positions and signs of the vortices are overlaid, and are also produced using only the ground truth density $\rho$.}
\end{figure*}

\begin{figure*}[t]
  \centering
  \includegraphics[width=\textwidth]{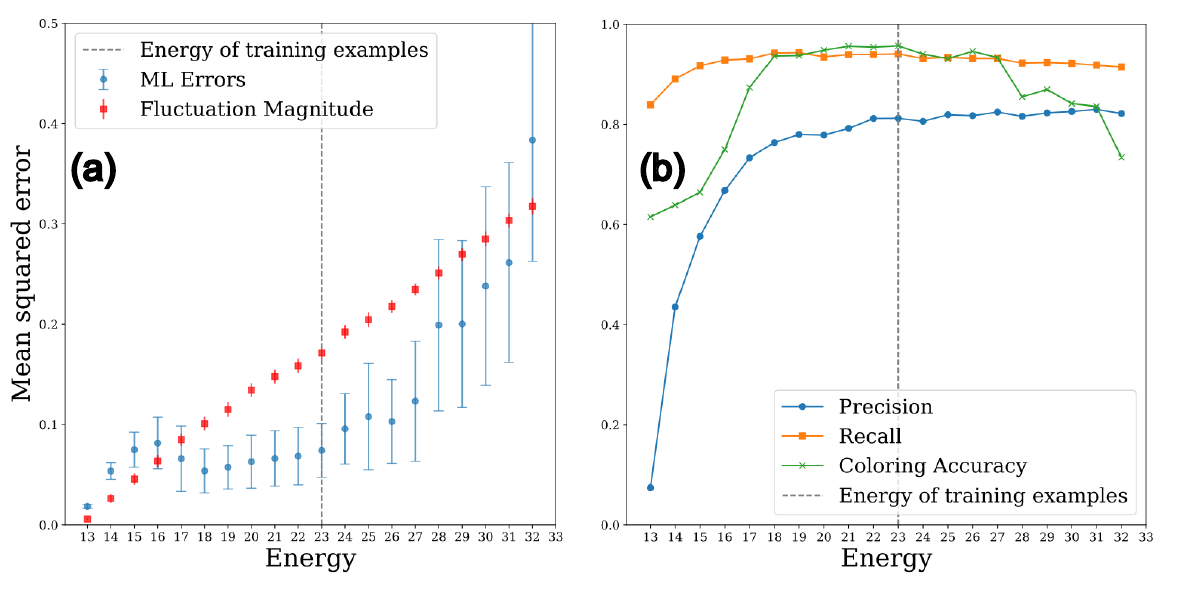}
  \caption{\label{fig:prediction_stats_with_background} Performance of our algorithm on the test set, which includes energies outside of the training set. (a) Error when predicting  $\nabla\phi$, calculated as a component and pixel-wise mean squared error within the masked region of each test example. The average value of $|\nabla\phi|^2$ within the masked region is plotted as baseline reference. (b) Performance of ML predicted vortex positions and signs. Precision, recall, and coloring accuracy of vortex predictions are shown, allowing up to a 1 pixel offset from the ground truth.}
\end{figure*}

In Fig. \ref{fig:current_predictions_with_background}, we show both the predictions of $\nabla\phi$ and vortex positions/colorings predictions for an example in the test set at $E = 23$. Since $\nabla\phi$ necessarily blows up near vortices, we instead visualize the predictions as the current field, $\rho\nabla\phi$, while emphasizing that the model predicted $\nabla\phi$ directly. As $\rho$ drops near the vortex cores, this allows us to visualize the predicted flows around vortices. We see that the model produces the flow of $\nabla\phi$ quite well, while underpredicting the magnitude of the gradient in some areas. For vortices, we see that almost all of the ground truth vortices and their colors are correctly predicted by the model; the model does, however, add spurious vortices in several places where there are no actual vortices. 

A quantitative analysis of the performance of the model is shown in Fig. \ref{fig:prediction_stats_with_background}. For each example in the test set, we find the pixelwise mean squared error between predicted $\nabla\phi$ and the ground truth $\nabla\phi$, averaged across pixels within a certain radius of the center of the trap, where the density is non-negligible - which we call the \textit{masked} area. We see from Fig. \ref{fig:prediction_stats_with_background} that error between the ground truth and the predicted values is low relative to the average value of $|\nabla\phi|^2$ for a wide range of energies around the training energy. 

To analyze the performance of the vortex identification, predicted vortices are matched one-to-one with ground-truth vortices if they are within 1 pixel of each other. This reflects an inherent deficiency in numerical vortex identification in the ground truth, where numerically detected vortices may be a pixel away from the actual vortex location. We use what we call \textit{off-by-one-recall} and \textit{off-by-one-precision} within the masked areas to quantify the accuracy of vortex localization, where:
\begin{equation}
\text{off-by-one-recall} = \frac{\text{\# Ground Truth-ML Matches}}{\text{\# Ground Truth Vortices}}
\end{equation}
\begin{equation}
\text{off-by-one-precision} = \frac{\text{\# Ground Truth-ML Matches}}{\text{\# Predicted Vortices}}
\end{equation}
Thus, if the model fails to predict the existence of a vortex near an actual vortex, the recall is lowered; if the model predicts the existence of a vortex where there is no actual vortex nearby, or predicts that many vortices exist near a single ground truth vortex, the precision decreases. We see from Fig. \ref{fig:prediction_stats_with_background} that the recall is ~0.9 for a wide range of energies around the training energy, and that precision is ~0.8. Finally, we quantify the vortex sign prediction accuracy by:
\begin{equation}
\text{Coloring Accuracy} = \frac{\text{\# Matches with Same Color}}{\text{\# Matches}}
\end{equation}
And we find that the coloring accuracy is ~0.9 for a wide range of energies around the training energy, indicating that the model can assign correct relative signs to vortices in many cases, given only the density as input.

\section{Algorithm Methodology \label{sec:algorithm}}
\begin{figure}[h]
      \centering
      \includegraphics[width=\columnwidth]{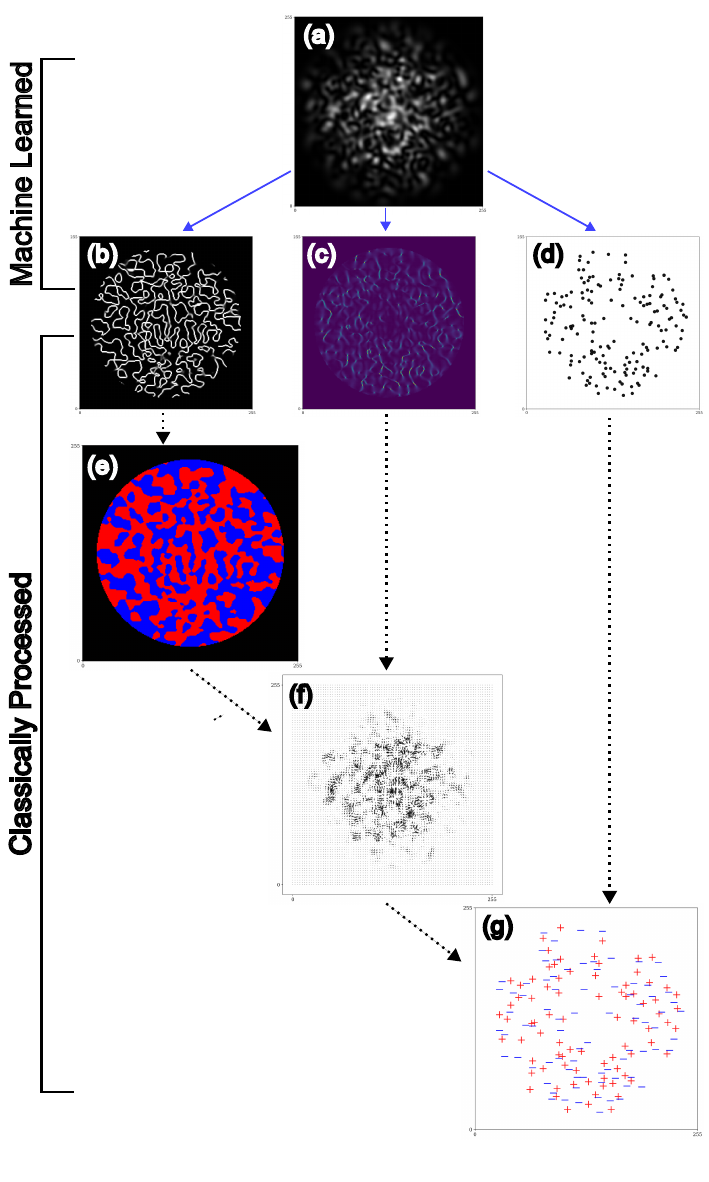}
      \caption{\label{fig:full_flow} A flow chart of the full algorithm. First, given the (a) input density, 3 separate deep learning models produce predictions for: (b) zeros of $\partial_{x}\phi$ and zeros of $\partial_{y}\phi$ (only the x field is shown here); (c) $|\partial_{x}\phi|$ and $|\partial_{y}\phi|$ (only the x field shown); and (d) predicted vortex positions. (e) A 2-coloring algorithm is then used to find signs of $\partial_{x}\phi$ and $\partial_{y}\phi$ separately. (f) The predicted signs, values of $|\partial_{x}\phi|$ and $|\partial_{y}\phi|$, and the quantization condition allow reconstruction of the full $\nabla\phi$. (g) Knowing $\nabla\phi$ then allows us to distinguish between vortices/antivortices.}
\end{figure}

We find that by using deep learning alone, we are unable to correctly predict the full gradient, even when allowing for a global sign error. Thus, our algorithm consists of a deep learning step to learn absolute values of the two components of the phase gradient, followed by a classical post-processing step to "color" the results and determine a consistent choice of signs, as shown in Fig. \ref{fig:full_flow}. 
    
 \subsection{Deep Learning}
The deep learning step relies on the U-Net architecture \cite{ronneberger_u-net_2015}. This architecture is used due to its power in image-to-image translation.
We use 3 separate deep learning models for 3 separate tasks, shown in Fig. \ref{fig:full_flow}b: 
\begin{enumerate}
    \item Prediction of the locations of the zeros of $\partial_x\phi$ and $\partial_y\phi$ - that is, where the derivatives change signs. If one imagines coloring the image based on the sign of $\partial_x\phi$, this corresponds to predicting the borders between regions of positive and negative signs (and equivalently for $\partial_y\phi$). We found that using two chained U-Nets, with the first U-Net performing the intermediate task of $ln|\partial_x\phi|$ and the downstream U-Net then predicting the zeros of $\partial_x\phi$, gave the best results. Predictions were returned as values between 0 and 1 for each pixel, representing the machine-learned probability of the sign crossing being at that location. 

    \item Prediction of $|\partial_x\phi|$ and $|\partial_y\phi|$. This step uses a U-Net.

    \item Prediction of the positions of vortices, without predicting the signs of the vortices. The U-Net predictions were returned as values between 0 and 1 for each pixel, and we follow the procedure in \cite{ye_u-net_2023}, with thresholding and non-maximum suppression, to produce binary predictions for vortex locations. 
\end{enumerate}

\subsection{Classical Post-Processing}
\begin{figure*}[t]
  \centering
  \includegraphics[width=\textwidth]{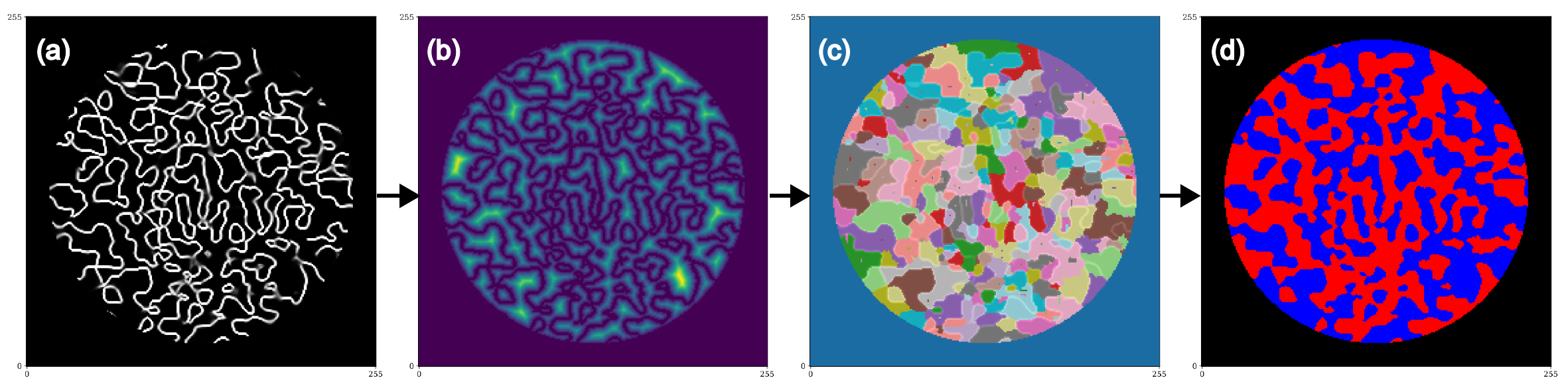}
  \caption{\label{fig:coloring_flow} An example flow of the 2-coloring algorithm to find the sign of $\partial_x\phi$ (with the same algorithm run separately for $\partial_y\phi$). (a) Noisy ML prediction of the boundaries between regions where $\partial_x\phi > 0$ and regions where $\partial_x\phi < 0$, returned as probabilities. (b) Distance transform from boundaries. (c) Watershed segmentation, seeded from the maxima in the distance transform. (d) Merging of watershed regions, accomplished by greedily minimizing an Ising-like functional defined by the ML predicted boundaries. }
\end{figure*}

\subsubsection{Sign Prediction from Boundaries}
After the deep learning stage, we have the ML-predicted borders, predicting where the sign of $\partial_x\phi$ changes across the borders (and similarly for $\partial_y\phi$). Given ground truth borders, assigning the sign of $\partial_x\phi$ to every pixel is a simple 2-coloring problem. However, the ML-predicted borders are returned as probabilities, and are not guaranteed to be always correct. The most involved part of the classical post-processing thus reduces to finding a configuration of the sign of $\partial_x\phi$ that maximally respects the ML-predicted borders; this process corresponds to the step between Fig. \ref{fig:full_flow}b and Fig. \ref{fig:full_flow}e. The full process is shown in Fig. \ref{fig:coloring_flow}.

The process consists of the following:

\begin{enumerate}

    \item Thresholding and distance transform. We apply a threshold to the ML-outputted probability map to create a binary boundary map, and compute distances from the closest boundaries to create a distance transformed image. 
    
    \item Watershed Segmentation and Graph Construction. Starting from the local maxima in the distance transformed image, we imagine filling those regions with water in a watershed transform, with the probabilities of the ML output giving the heights of the landscape. This produces an initial ``over-segmented'' image, where each region $\mathcal{R}_i$ is presumed to share the same sign. We treat this segmented image as a graph where each region is a node, and the predicted boundary probability $p_{ij}$ between two adjacent regions $\mathcal{R}_i$ and $\mathcal{R}_j$ acts as the weight of the connecting edge.

    \item Color Assignment. We assign a sign of $s_i = \pm 1$ to each node (region) in the constructed graph such that the assignment is maximally consistent with the boundary probabilities predicted by the neural network. First, we define (unnormalized) probabilities for regions sharing/not sharing signs based on the set of pixels $\partial \Omega_{ij}$ forming their boundary:
    \begin{equation}
    P(s_i = s_j) = \prod_{\mathbf{r} \in \partial \Omega_{ij}} (1 - p(\mathbf{r}))
    \end{equation}
    \begin{equation}
    P(s_i \ne s_j) = \prod_{\mathbf{r} \in \partial \Omega_{ij}} p(\mathbf{r})
    \end{equation}

    And then find the configuration $\{s_i\}$ that maximizes the joint probability:
     \begin{equation}
    P(\{s_i\}) = \prod_{\langle i, j \rangle} \frac{1}{2}\left[(1+s_is_j)P(s_i = s_j) + (1-s_is_j)P(s_i \ne s_j)\right]
    \end{equation}
    Where $\langle i, j \rangle$ are neighbors in the graph. Note that, upon taking the logarithm, this is equivalent to finding the ground state of an Ising model with energy:

    \begin{equation}
    E(\{s_i\}) = \sum_{\langle i, j \rangle} s_i s_j \ln \left( \frac{P(s_i \ne s_j)}{P(s_i = s_j)} \right)
    \end{equation}

    We solve this via a greedy algorithm detailed in the appendix. While other solutions (e.g simulated annealing) may produce better solutions in some cases, we find that the simple greedy algorithm reproduces ground truth signs very well in most cases. 
    \end{enumerate}

\subsubsection{Final Gradient and Vortex Predictions}

Having predicted the sign of $\partial_x\phi$ and the value of $|\partial_x\phi|$, the full $\partial_x\phi$ and $\partial_y\phi$ are each reconstructed up to a global sign. However, this does not fix the relative sign between $\partial_x\phi$ and $\partial_y\phi$; to obtain the relative sign, we try both $\nabla\phi = (\partial_x\phi, \partial_y\phi)$ and $\nabla\phi' = (\partial_x\phi, -\partial_y\phi)$, and numerically compute circulations around 2x2 loops (see Eq.\ref{quantization_eq}). We choose the choice that reproduces the circulation quantization condition more closely. 

Finally, we assign signs to vortices by numerically computing circulations around the border of a 3x3 block centered on each ML-predicted vortex position, using our predicted $\nabla\phi$, and assigning vortex charges based on the signs of the circulation. 

\section{Predictions with No Thermal Background}
\begin{figure*}[t]
  \centering
  \includegraphics[width=\textwidth]{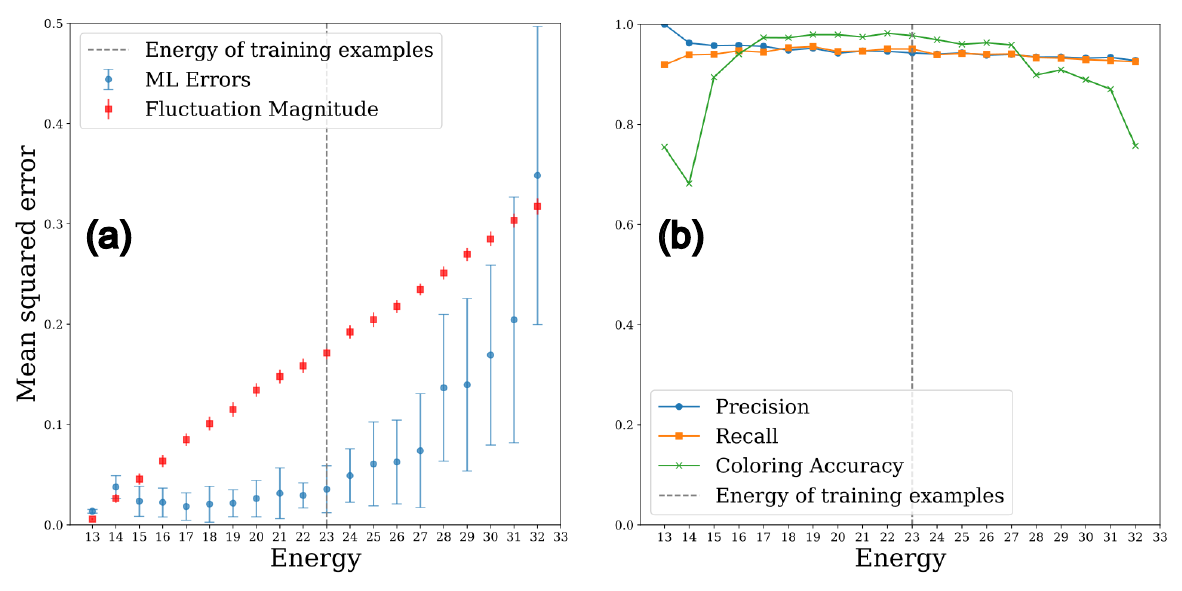}
  \caption{\label{fig:prediction_stats_without_background} Same as \ref{fig:prediction_stats_with_background}, but for our algorithm trained and tested on a dataset with no added thermal background.}
\end{figure*}
To further probe the utility of our algorithm, we apply it to a case where both the training set and the testing set include no thermal background density - that is, where the ML has direct access to the density of the classical field alone. As with the thermal background case, we find the mean squared error of the predicted $\nabla\phi$, as well as vortex core position and coloring errors, which we show in Fig.\ref{fig:prediction_stats_without_background}. While the trends are similar, we find that both the $\nabla\phi$ reconstruction errors and the vortex reconstruction errors are significantly lower.

The jump in precision of the vortex reconstruction is very noticeable. We hypothesize that the lack of precision in the case with thermal background is due to the varying strength of the thermal background across energies; as the thermal background density means that vortex cores no longer have 0 density, the best that the U-Net can do is to find relative density fluctuations resembling a vortex core - which can sometimes be the case even when there is no vortex core there. When there is no thermal background density, the vortex core actually has 0 density, in addition to having the relative density fluctuation profile of a vortex core - allowing the ML to accurately distinguish the vortex core from other density fluctuations.

\section{Conclusion \label{sec:conclusion}}

This work presents an end-to-end machine learning framework combined with classical post-processing techniques for solving the problem of reconstructing the phase field and classifying topological defects in a two-dimensional BEC from in-situ density images alone. We use the image processing power of the U-Net to predict a time-reversal-invariant representation of the magnitude of the components the phase gradient, as well as to find the location of vortex positions. Using classical post-processing, we fully reconstruct the gradient of the phase up to a global sign, and assign charges to vortices, a capability previously limited in methods relying solely on density information. For both cases with and without a thermal background, we find ground truth vortices and color them in correctly with high ($>$90\%) accuracy. For the case without a thermal background, we find that the precision of vortex core location is significantly improved over the thermal background case.

The framework introduced here is immediately relevant for the direct analysis of experimental density snapshots of boson systems. From a very general point of view, the problem addressed here is the learning of signed quantities from measurements of a scalar quantity. The combination of image translation to learn scalar structures and deterministic post-processing to assign signs may be powerful in other contexts.

An important direction for future research is to determine whether the approach presented here can be generalized to enable the inference of quantal correlations in other systems.  We note that in our work each snapshot has a definite value of the phase field uniquely related to a definite density. In the true quantum case one wishes to learn the expectation value  of an operator field in a  wave function from a set of projections of this wave function onto a particular (for example density) basis, so generating a set of density fields all corresponding to the same expectation value. This issue can be overcome by analysing an ensemble of experimental images. A more severe potential difficulty is that  the operator field whose expectation value is sought may not correspond to a particular feature in any of the images, rendering the U-Net approach used here less powerful. Investigation of other ML architectures is important.

\begin{acknowledgments}
We thank S. Will, A. Sengupta, A. Georges, C. Roth, and A. Valenti for advice and helpful conversations. The work of J.L was supported by the U.S. Department of
Energy, Office of Science, Office of Advanced Scientific Computing Research, Department of
Energy Computational Science Graduate Fellowship under Award Number DE-SC0025528. The work of AJM was supported in part by Programmable Quantum Materials, an Energy Frontier Research Center funded by the US Department of Energy, Office of Science, Basic Energy Sciences, under award no. DE-SC0019443. The Flatiron Institute is a division of the Simons Foundation.
\end{acknowledgments}

\textbf{Disclaimer:}
This report was prepared as an account of work sponsored by an agency of the
United States Government. Neither the United States Government nor any agency thereof, nor
any of their employees, makes any warranty, express or implied, or assumes any legal liability or responsibility for the accuracy, completeness, or usefulness of any information, apparatus, product, or process disclosed, or represents that its use would not infringe privately owned rights. Reference herein to any specific commercial product, process, or service by trade name, trademark, manufacturer, or otherwise does not necessarily constitute or imply its endorsement, recommendation, or favoring by the United States Government or any agency thereof. The views and opinions of authors expressed herein do not necessarily state or reflect those of the United States Government or any agency thereof.

\appendix
\section{PGPE Simulations}
\label{app:PGPE Simulations}


In this appendix we outline the derivation of the dimensionless 2D projected Gross--Pitaevskii equation (PGPE) from its dimensionful form, and summarize how the equation is time-evolved in a  harmonic trap and sampled for our dataset. We follow \cite{dion_spectral_2003} and \cite{blakie_projected_2005}, adapted to our 2D case. 
\subsection{Projected Gross--Pitaevskii equation}

We consider a two-dimensional projected Gross--Pitaevskii equation (PGPE)
describing the dynamics of a classical Bose field
$\Psi(\mathbf{r},t)$ confined in a harmonic trap. The dimensionful PGPE is
\begin{equation}
i \hbar \frac{\partial \Psi}{\partial t}
=
\left(
-\frac{\hbar^2}{2m}\nabla^2
+ V_{\mathrm{trap}}(\mathbf{r})
\right)\Psi
+
\mathcal{P}\!\left\{
\frac{4\pi\hbar^2 a}{m}
\,|\Psi|^2\Psi
\right\},
\label{eq:pgpe_dimensional}
\end{equation}

where $\nabla^2$ is the two-dimensional Laplacian, $V_{\mathrm{trap}}(\mathbf{r}) = \tfrac{1}{2}m\omega^2(x^2+y^2)$ is the harmonic trapping potential, $a$ the s-wave scattering length, and $\mathcal{P}$ denotes a projector restricting the dynamics to a finite set of low-energy single-particle modes defining the coherent region. Note that here the classical field is normalized as:
\begin{equation}
\int d^2\mathbf{r}\,|\Psi|^2 = N_c,
\end{equation}
Where $N_c$ is the number of particles in the classical region. 

\subsection{Dimensionless formulation}

We nondimensionalize Eq.~\eqref{eq:pgpe_dimensional} using the harmonic trap
frequency $\omega$ to set the characteristic scales
\begin{equation}
x_0 = \sqrt{\frac{\hbar}{m\omega}}, \qquad
t_0 = \omega^{-1}, \qquad
E_0 = \hbar\omega .
\end{equation}
Introducing dimensionless coordinates and time
$\tilde{\mathbf{r}} = \mathbf{r}/x_0$ and $\tilde{t} = t/t_0$, and writing the
classical field as
\begin{equation}
\psi(\mathbf{r},t)
=
\sqrt{N_c}\,x_0^{-1}\,\tilde{\psi}(\tilde{\mathbf{r}},\tilde{t}),
\end{equation}
where $\tilde{\psi}$ is normalized to unity,
\begin{equation}
\int d^2\tilde{\mathbf{r}}\,|\tilde{\psi}(\tilde{\mathbf{r}},\tilde{t})|^2 = 1,
\end{equation}
yields the dimensionless PGPE
\begin{equation}
i \frac{\partial \tilde{\psi}}{\partial \tilde{t}}
=
\left[
-\frac{1}{2}\tilde{\nabla}^2
+ \frac{1}{2}(\tilde{x}^2+\tilde{y}^2)
\right]\tilde{\psi}
+
\mathcal{P}\!\left\{
C_{\mathrm{nl}}\,|\tilde{\psi}|^2\tilde{\psi}
\right\},
\label{eq:pgpe_2d_dimensionless}
\end{equation}
where the dimensionless nonlinear coupling constant is
\begin{equation}
C_{\mathrm{nl}} = 4\pi a N_c\sqrt\frac{m\omega}{\hbar}
\end{equation}
And all the physical parameters of the system are thus captured by $C_{\mathrm{nl}}$. 

\subsection{Time evolution and equilibration protocol}

The dimensionless PGPE, Eq.~\eqref{eq:pgpe_2d_dimensionless}, is evolved
numerically using a spectral representation in the eigenbasis of the
two-dimensional harmonic oscillator. The field is expanded as
\begin{equation}
\tilde{\psi}(\tilde{x},\tilde{y},\tilde{t})
=
\sum_{n_x,n_y\in\mathcal{C}}
c_{n_x n_y}(\tilde{t})\,
\phi_{n_x}(\tilde{x})\phi_{n_y}(\tilde{y}),
\label{eq:ho_expansion_2d}
\end{equation}
where $\phi_n(\tilde{x})$ are the normalized eigenfunctions of the
one-dimensional harmonic oscillator,
\begin{equation}
\left(
-\frac{1}{2}\frac{d^2}{d\tilde{x}^2}
+
\frac{1}{2}\tilde{x}^2
\right)\phi_n
=
\epsilon_n \phi_n,
\qquad
\epsilon_n = n + \frac{1}{2},
\end{equation}
and the coherent region $\mathcal{C}$ is defined by the energy cutoff
\begin{equation}
\mathcal{C}
=
\left\{
(n_x,n_y)\,:\,
\epsilon_{n_x} + \epsilon_{n_y} \le E_{\mathrm{cut}}
\right\}.
\end{equation}

Substituting Eq.~\eqref{eq:ho_expansion_2d} into the PGPE and projecting onto a
basis state yields equations of motion for the mode amplitudes,
\begin{equation}
\frac{d c_{n_x n_y}}{d \tilde{t}}
=
-i(\epsilon_{n_x}+\epsilon_{n_y})c_{n_x n_y}
-
i C_{\mathrm{nl}}\,F_{n_x n_y}[\tilde{\psi}],
\label{eq:coeff_eom_2d}
\end{equation}
with nonlinear coupling
\begin{equation}
F_{n_x n_y}[\tilde{\psi}]
=
\int d\tilde{x}\,d\tilde{y}\,
\phi_{n_x}(\tilde{x})\phi_{n_y}(\tilde{y})
\,|\tilde{\psi}|^2\tilde{\psi}.
\end{equation}

The nonlinear coupling is efficiently evaluated by a Gauss-Hermite quadrature - for more details, see \cite{dion_spectral_2003}. We then time-evolve an initial state using 4th order Runge-Kutta iteration. 
\subsection{Simulation parameters}

All simulations reported here use parameters appropriate for
${}^{87}\mathrm{Rb}$ atoms, with mass
$m = 86.9\,\mathrm{amu}$ and $s$-wave scattering length
$a = 106\,a_0$, where $a_0$ is the Bohr radius. The harmonic trap frequency is
$\omega = 200\pi\,\mathrm{Hz}$, and the number of particles in the coherent
region is fixed to $N_c = 1.5\times10^4$.

For each target energy, the ground state was first obtained by imaginary-time evolution of the
PGPE. Controlled perturbations were then applied by adding random complex
amplitudes to the harmonic-oscillator coefficients, with magnitudes drawn from
a Gaussian distribution and phases chosen uniformly on $[0,2\pi)$. The overall
perturbation strength was adjusted to achieve the desired total energy.

The number of modes retained in the coherent region was energy dependent and
chosen such that the least occupied mode at the end of the evolution contained
at least one particle and fewer than ten particles, ensuring consistency with
the classical-field approximation \cite{blakie_dynamics_2008}. This criterion resulted in cutoffs ranging
from $E_{\mathrm{cut}}=20$ (190 modes) at energy $E=13$ to
$E_{\mathrm{cut}}=48$ (1128 modes) at energy $E=32$.

Time evolution was carried out for $200$ trap periods for test datasets, with
$100$ uniformly spaced samples taken from the final $100$ periods after
equilibration. For training and validation datasets, the field was evolved for
$400$ trap periods, and $1500$ uniformly spaced samples were collected from the
final $300$ periods. In all cases, observables and datasets were constructed
from these late-time samples, corresponding to equilibrium configurations of
the microcanonical PGPE dynamics.

\section{U-Net Architecture and Training} 
\begin{figure*}[t]
  \centering
  \includegraphics[width=\textwidth]{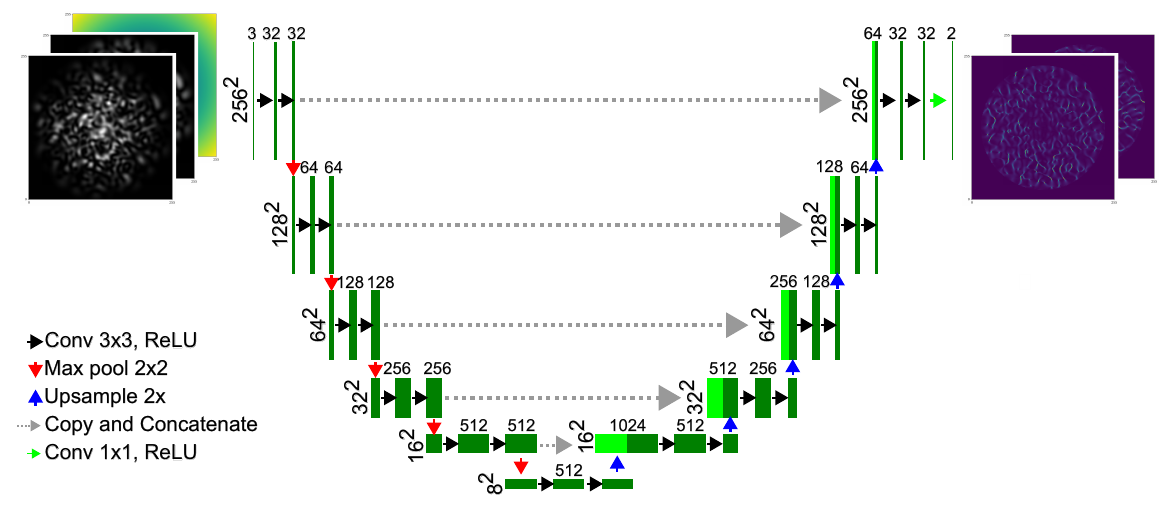}
  \caption{\label{fig:unet} The U-Net architecture for the problem of predicting $|\partial_x\phi|$ and $|\partial_y\phi|$, with channel widths $[3, 32, 64, 128, 256, 512]$. On the left, we input 3 256x256 images - the density, the log-transformed density, and the confining potential (which is the same across all examples). At each stage of the U-Net, the channel dimension is represented by the number on top, while the image height/width are represented by the number to the left; for instance, the input in the first layer is 3x256x256, before being transformed via convolution into a 32x256x256 block. The output is 2x256x256, with the 256x256 slices being $|\partial_x\phi|$ and $|\partial_y\phi|$, respectively.}
\end{figure*}
U-Nets are a class of convolutional neural networks designed for learning mappings between input and output fields while preserving spatial structure. The architecture consists of a contracting (encoder) path and an expanding (decoder) path arranged symmetrically in a “U” shape. In the encoder, successive convolution and downsampling operations extract increasingly abstract, coarse-grained features while reducing spatial resolution. In the decoder, upsampling and convolution progressively restore spatial resolution. Crucially, skip connections link corresponding layers in the encoder and decoder, allowing high-resolution feature maps from early layers to be concatenated with upsampled features. These skip connections enable the network to combine global contextual information with local spatial detail, improving reconstruction accuracy and gradient flow during training. As a result, U-Nets are particularly well suited for tasks involving structured, spatially localized outputs, such as image segmentation or learning field-to-field mappings.

All neural networks were trained using the Adam optimizer with task-specific
learning rates \cite{kingma_adam_2017}. The inputs to all models were three-channel images consisting
of the condensate density, the log-transformed density, and the external
potential. In the harmonic trap considered here, the potential is a known,
fixed function of position and effectively encodes the radial distance of each
pixel from the trap center. Including this channel provides the network with
explicit geometric context.

For the density-to-phase-gradient task, a single U-Net with channel widths
$[3, 32, 64, 128, 256, 512]$ was used to predict the two components of the
absolute phase gradient. The network was trained for $30$ epochs with batch
size $8$ and learning rate $10^{-3}$ using a masked mean-squared error loss,
where the mask restricts the loss to physically relevant regions of the field.
For vortex detection, an identical U-Net architecture was employed, but with a
single output channel corresponding to vortex locations. This model was trained
for $30$ epochs with batch size $8$ and learning rate $5\times10^{-4}$ using a
binary cross-entropy loss weighted by class frequency to account for the strong
imbalance between vortex and non-vortex pixels.

For boundary detection, a two-stage chained U-Net architecture was used. The
first network, with channel widths $[3, 32, 64, 128, 256, 512]$, predicts two
intermediate channels corresponding to the logarithm of the magnitude of the
phase gradient. This output is transformed and passed as input to a second
U-Net with channel widths $[2, 32, 64, 128, 256, 512]$, which predicts two final
channels encoding the locations of phase sign boundaries. Both intermediate and
final outputs are standardized by normalizing with their dataset mean and
variance. Training was performed for $30$ epochs with batch size $8$ and
learning rate $10^{-3}$ using a masked mean-squared error loss computed over all
four output channels, while inference and evaluation were carried out using only
the two final boundary channels.

\section{Greedy Likelihood Maximization for 2-Coloring}

Since finding the true ground state of the Ising model is computationally intensive, we approximate the minimization using a heuristic, greedy algorithm inspired by Dijkstra's shortest path algorithm. This method starts by selecting a the largest region (in terms of number of pixels) as the starting region. It then iteratively "colors in" neighboring regions, choosing the sign assignment ($s_i = \pm 1$) for each new region that minimizes the accumulated energy penalty based on the decisions made so far. To prevent settling into poor local minima, this entire process is repeated with multiple random restarts (e.g., $\mathbf{20}$ restarts) using different initial regions. The final sign assignment is taken from the iteration with the lowest overall total energy.


\bibliography{bec_ml_refs}

\end{document}